\documentstyle[preprint,aps,eqsecnum]{revtex}
\begin{document}
\preprint{\baselineskip 18pt{\vbox{\hbox{SU-4240-590} \hbox{December, 1994}}}}
\vspace{35mm}
\title{Edge States in Gravity and Black Hole Physics}
\vspace{15mm}
\author{A.P. Balachandran, L. Chandar,  Arshad Momen}
\vspace{15mm}
\address{Department of Physics, Syracuse University,\\
Syracuse, NY 13244-1130, U.S.A.}
\vspace{15mm}
\maketitle
\begin{abstract}
We show in the context of Einstein gravity that
the removal of a spatial region leads to the
appearance of an infinite set of observables and their associated edge states
localized at its boundary. Such a boundary occurs in certain approaches to the
physics of black holes like the one based on the membrane paradigm. The edge
states can contribute to black hole entropy in these models.
A ``complementarity principle" is also shown to emerge whereby certain ``edge"
observables are accessible only to certain observers. The physical significance
of edge observables and their states is discussed using their similarities to
the corresponding quantities in the quantum Hall effect. The coupling of the
edge states to the bulk gravitational field is demonstrated in the context of
(2+1) dimensional gravity.
\end{abstract}
\vspace{5mm}
\pagebreak
\vfill

\newcommand{\be}{\begin{equation}}
\newcommand{\ee}{\end{equation}}
\newcommand{\bea}{\begin{eqnarray}}
\newcommand{\eea}{\end{eqnarray}}
\newcommand{\real}{{\rm l}\! {\rm R}}
\newcommand{\ra}{\rightarrow}
\newcommand{\tr}{{\rm tr}\;}
\section{Introduction}

Since the discovery of black hole radiation by Hawking \cite{haw1,andy}, the
eventual fate of information falling into a black hole has remained an
unsolved puzzle. The different
scenarios that have been proposed can be
summarized in the following two classes:

\qquad  (1) Information going into a black hole is irretrievably
lost so that unitary quantum theory breaks down \cite{haw2}.

\qquad  (2) Information that was thought to be ``lost'' reappears in
some form
thus saving conventional quantum physics \cite{thooft}.

Quantum black hole physics is usually studied in a semiclassical
approximation.  Typically it is shown
that processes like Hawking radiation and the scattering
of matter fields by black holes are governed by physics occurring close to
the
event horizon \cite{susskind}.

There have been several interesting proposals
\cite{susskind,membranebook,maggiore} to treat the black hole horizon as a
membrane
 with dynamical degrees
of freedom attached to it. It is found that for {\it any}
external stationary
observer,  the black hole
can be well approximated
by hypothesizing a stretched membrane having certain classical properties,
surrounding the hole at a
small distance
outside its event horizon. A physical justification
for this procedure is that particles
classically can never leave the interior of the black hole  or reach it from
outside in a
finite time according to any such observer.
 Thus it seems that any such person can
study the quantization of the system after removing the interior of
the black hole and replacing it by a membrane. The manifold in this approach
thus has a boundary.

In this paper, we study the quantum physics of black holes for such observers,
or more
generally features of quantum  gravity on manifolds with spatial boundaries. We
show that the presence of the boundary  necessarily leads to an
{\it infinite} set of {\it
observables} which are completely
{\it localized at this boundary} in the absence of anomalies. [The important
issue of anomalies and their significance is also addressed in this paper.]
These are obtained here in analogy
to  ``edge'' observables
in gauge theories defined on manifolds with boundaries \cite{sei,bal}.
Such observables, defined in the context of gauge theories, have enjoyed
good physical interpretation in many examples of condensed matter physics.
For instance, it has been known for some time that many of the essential
features of  the fractional quantum Hall effect (FQHE) are captured purely
by the
existence of these edge observables\cite{duality}.  We will discuss more
about this analogy
with the quantum Hall effect (QHE) later on in this paper.
{}From previous work dealing with the construction of these observables for
pure
gauge theories \cite{bal} and also from the construction exhibited in
this paper, one realizes that the edge
observables are independent of observables defined in the bulk and that
they commute with the Hamiltonian.  This is quite interesting because it means
that, no matter what the quantum theory of the bulk observables is, there is
always a Hilbert space living entirely on the edge.
One can then talk of excitations which only involve the edge  and leave
the bulk unaffected  in the absence of any coupling between them.

Hence, in spite  of the deep problems in quantizing gravity, one can still
examine the relatively simpler task of quantization of the gravitational edge
observables and  the associated edge states at the
boundary ( which, for a
black hole, is $S^2 \times \real$).
In this way we can also hope to make meaningful physical predictions
because of the above mentioned claim that the quantum aspects of a black hole
are linked to the existence of an imaginary boundary near the event horizon.

It merits emphasis that the stretched membrane \cite{membranebook} is
to be
thought of as surrounding the black hole at a small but finite distance
outside its event horizon. For this reason, our boundary will also
assumed
to be situated
slightly away from the actual event horizon. There is also a good technical
reason for
this assumption since the induced metric becomes degenerate on the event
horizon.

It is worth pointing out one other important observation emerging from
the analysis
of Section 3.  We notice the fact that even though there
exists an infinite set of observables at the edge of the black hole,
not {\it all of these} are relevant for {\it all the observers}.
In fact, a sort of
complementarity principle can be shown
according to which the existence of certain observables
as ``good'' observables will preclude
some others from being so.  The choice of the ``good"
observables is dictated by
physical considerations.  Thus, for
example, we shall see that for an asymptotic observer, physics will dictate
that there exists an infinite sub-family of ``good"  edge observables
whereas for an observer stationed
close to the
event horizon, only a finite number
out of this infinite set survive as
``good'' observables.  This difference between the notion of ``good''
observables for these two observers certainly needs to be understood more
carefully since this may be at the heart of many of the conceptual difficulties
concerning the quantum physics of black holes.  [ In this connection, see
\cite{susskind}, where a different kind of complementarity principle has been
discussed. ]

The organization of the paper is as follows.  In Section 2, we briefly review
the canonical formulation of pure gravity in (3+1) dimensions for the
convenience of the reader, following the
approach of \cite{ash} closely.  In Section 3, we use this formalism in (3+1)
dimensions for a manifold
obtained by removing a spatial ball. In black hole physics, this ball may be
regarded as enclosing the hole including its event horizon in its interior,
its boundary being the membrane.  We show that,
in such a spacetime, we are
naturally led to certain observables that are confined purely to the edge.
In Section 4, we discuss the
analogy of the aforementioned edge observables with similar observables arising
in condensed matter physics, notably in the QHE.  In particular, we argue that
even though formally the edge observables do not mix with the other observables
living in the bulk, this ceases to be the case especially when we have an {\it
anomalous} coupling between the edge and the bulk.  Such a coupling is not a
matter of choice, it being required from very general arguments of gauge
invariance in the case of the QHE \cite{duality}.
 The surprise is that this analogy
extends even to the case of gravity, where it turns out that an anomalous
coupling between the edge and the exterior is forced on us if we require
general coordinate invariance (diffeomorphism invariance).
In Section 5, we explicitly demonstrate that such a
coupling indeed appears when we are dealing with (2+1) gravity.  The final
Section 6 outlines an argument suggesting that the edge and bulk states couple
in (3+1) dimensions as well. This argument is unfortunately incomplete as we
lack a satisfactory description of edge dynamics.

\section{Canonical Formulation of the Einstein-Hilbert Action}

Consider a four-manifold $M$ which is topologically $\Sigma \times \real$
and let its time-slices $\Sigma_{t}$ be parametrized by
$t$. Hereafter we assume that $\Sigma_t$ is diffeomorphic to the exterior of a
ball ${\cal B}_3$ in  ${\real}^3$.   Let
$t^{a}$ be the vector field whose affine parameter is $t$ and let $n_{a}$
denote
the unit normal to the surfaces $\Sigma_{t}$ in the direction of increasing
$t$.  A metric $g_{ab}$ on $M$ induces a metric $q_{ab}$ on $\Sigma_{t}$.
Since $\Sigma_{t}$ is the spatial slice at time $t$, we need
$q_{ab}$ to have +++ signature. Then $n_{a}$, being normal to $\Sigma_{t}$,
will be timelike and (say) future-directed.  Thus we have the relation
\be
q_{ab}=g_{ab}+ n_{a}n_{b}. \label{1}
\ee
It gives
\be
g_{ab}t^{b}=N_{a} + \Gamma n_{a}, \label{2}
\ee
$N^a$ being tangent to $\Sigma_t$.  Here $\Gamma$ and $N^a$ are commonly
referred  as the `lapse' function and the `shift' vector field
respectively \cite{ash}.

Since we want to interpret $t^{a}$ as the vector field along which
the
variables defined on $\Sigma_{t}$ evolve, it has to be timelike and
future-directed.  Thus
\begin{eqnarray}
\Gamma ^{2} - N^{a}N_{a} >0, \nonumber \\
\Gamma >0. \label{3}
\end{eqnarray}
Furthermore, since we are interested in asymptotically flat spacetimes, the
appropriate asymptotic conditions to impose on the
lapse and  shift, in order that $t^{a}$ reduces to a timelike Killing
vector field normal to the spatial slice at spatial infinity, are
\be
\Gamma \;\; \rightarrow \;\; 1,\;\;\;\; N_{a} \;\; \rightarrow \;\; 0
\label{lapshi}
\ee

The Einstein-Hilbert action is
\be
S = \int (-g)^{\frac{1}{2}}\; ^{(4)}R \; d^{4}x, \label{EH}
\ee
where $g$ is the determinant of $g_{ab}$ and $^{(4)}R$ is the Ricci scalar of
the four-manifold.
It gives the Lagrangian
\be
L' = \int d^{3}x\; (-g)^{\frac{1}{2}}\; ^{(4)}R. \label{5}
\ee
at time $t$.
Here ($x_{1}, x_{2}, x_{3}$) are coordinates on $\Sigma _{t}$ which we
identify hereafter with $\Sigma$.

Now {\it apart from surface terms}, this $L'$ is the same as
\be
L = \int _{\Sigma} d^{3}x\; q^{\frac{1}{2}}\Gamma
({^{(3)}R} + K^{ab}K_{ab} - K^{2}),\label{6}
\ee
where $q$ is the determinant of $q_{ab}$, $^{(3)}R$ is its Ricci scalar and

\bea
K_{ab} &=& \frac{1}{2\Gamma }({\cal L}_{t}q_{ab} - {\cal L}_{N}q_{ab}),
\label{extrinsic} \nonumber\\
K&=&q^{ab}K_{ab}= K^a_a
\eea
The advantage of using (\ref{6}) over (\ref{5}) is that (\ref{6}) is a
functional only of $q_{ab}$ and $\dot{q}_{ab} (={\cal L}_{t}q_{ab})$ whereas
(\ref{5}) depends also on $\ddot{q}_{ab}$.

The  Lagrangian $L$ gives the following expression for the conjugate momentum
$p^{ab}$ :
\be
p^{ab} = \frac{\delta L}{\delta \dot{q}_{ab}}= \sqrt{q}(K^{ab} - Kq^{ab}).
\label{7}
\ee

We can now do a Legendre transform to go to the Hamiltonian.  Since $L$ does
not contain terms with $\dot{\Gamma}$ and $\dot{N}^{a}$, the corresponding
momenta
are constrained to be zero. They also generate secondary constraints.
The Hamiltonian (again up to surface terms) and the secondary constraints are
\begin{eqnarray}
H' &=& \int _{\Sigma} d^{3}x\; p^{ab}\dot{q}_{ab} - L\nonumber \\
  &=& \int _{\Sigma} d^{3}x\; \Gamma [-q^{\frac{1}{2}} \;^{(3)} R +
q^{-\frac{1}{2}}(p^{ab}p_{ab}
-\frac{p^{2}}{2})] + \int _{\Sigma} d^{3}x\; p^{ab}{\cal
L}_{N}q_{ab},\label{8}
\eea
\bea
D_{a}p^{ab} &\approx &   0 \label{extra} \\
-{q^{\frac{1}{2}}}\;{^{(3)}R} + q^{-\frac{1}{2}}(p^{ab}p_{ab} -\frac{p^{2}}{2})
&\approx &  0
\label{9}
\eea

 The primary constraints have vanishing Poisson brackets (PB's)
with (\ref{extra}),(\ref {9})
which also do not contain $\Gamma$ and $N^a$. The former can therefore be
ignored. The secondary constraints do not generate further constraints,
their PB's
with the Hamiltonian being weakly zero.  They are also first class.

In evaluating the Poisson brackets
of the constraints amongst themselves and for finding their action on the phase
space it is of course necessary to smear them with test functions so that
they become differentiable \cite{regge}.  The vector constraint is to be
smeared
with a form
$V_{a}$ that vanishes  at the boundaries  of the manifold and the scalar
constraint is to be  smeared with a test function $S$ that vanishes (along with
derivatives) at the
boundaries. The boundaries here are the boundary $\partial {\cal B}_3$ of
${\cal B}_3$ and
spatial infinity.  The smeared constraints are
\begin{eqnarray}
{\cal V}_{V}(q,p) & = & - 2 \int _{\Sigma} d^{3}x\; V_{a}D_{b}p^{ab} \approx 0,
\label{9.5}\\
{\cal S}_{S}(q,p) & = & \int _{\Sigma} d^{3}x\; S[-{q^{\frac{1}{2}}}\;{^{(3)}R}
+q^{-\frac{1}{2}}(p^{ab}p_{ab}-\frac{p^{2}}{2})] \approx 0, \label{10}
\eea
where
\bea
V_a |_{\partial \Sigma} = 0 \label{bc1} \\
S |_{\partial \Sigma} = 0, \qquad D_a S|_{\partial \Sigma}=0  \label{bc2}
\end{eqnarray}
The above conditions on the form $V_{a}$ and the function $S$ follow purely
from requiring differentiability in the phase space variables $q_{ab}$ and
$p^{ab}$ of (\ref{9.5}) and (\ref{10}).

The PB's  among the constraints are
\begin{eqnarray}
\{ {\cal V}_{V_{1}}\;,\;{\cal V}_{V_{2}}\} & = & {\cal V}_{[V_{1}\;,\;V_{2}]},
\nonumber \\
\{ {\cal V}_{V}\;,\;{\cal S}_{S}\} & = & {\cal S}_{{\cal L}_{V}S},\nonumber \\
\{ {\cal S}_{S_{1}}\;,\;{\cal S}_{S_{2}}\} & = & {\cal
V}_{S_{1}D\,S_{2}-S_{2}D\,S_{1}}. \label{11}
\end{eqnarray}

Clearly, the advantage that (\ref{6}) had over (\ref{5}) would be maintained
if we add a surface term to (\ref{6}) which is a functional only of $q_{ab}$
and not of $\dot{q}_{ab}$ \cite{regge}.
In fact, (\ref{8}) is meaningless as it stands because the boundary condition
(\ref{lapshi}) is different from those required of the test functions $S$
in (\ref{bc2}). Thus we need to add suitable surface terms to make the
Hamiltonian, for example, differentiable.

 If the Hamiltonian $H$ obtained after including these surface
terms is to describe evolution for an asymptotic observer, $\Gamma$ and $N_{a}$
have to satisfy the asymptotic conditions specified in (\ref{lapshi}) while
both of them have to vanish at the boundary surrounding the black hole
interior.  This is because in
the reference frame of the asymptotic observer, the observer's own time runs at
unit rate while the time on the horizon of the black hole has ``stopped''. We
can thus write, using  (\ref{9.5}),
\bea
H= \int_\Sigma d^3x \;\Gamma \left[ -{q^{\frac{1}{2}}}\;{^{(3)}R}
+ q^{-\frac{1}{2}}\left(
p^{ab}p_{ab}- \frac{p^2}{2}\right) \right] + ( {\rm surface\;\; terms}),
\label{new} \\
\Gamma|_{\partial \Sigma} = 0 ;\qquad \;\; \Gamma|_{spatial \;\;infinity} = 1
\label{bc3}
\eea
where the surface terms are to be
chosen so that $H$ becomes differentiable in
$q_{ab}$ and $p^{ab}$. We will further discuss these terms below.

\section{Observables Living at the Edge}

The construction of edge observables uses a trick that we have
already employed in the previous Section.  It is illustrated by the distinction
between the test functions $\Gamma$ and $S$ for the
Hamiltonian (\ref{new}) and the  smeared constraint (\ref{10}).  Since $H$
becomes a first class constraint if $\Gamma$ goes to zero  at the boundaries,
we can regard $H$ as localized at the boundaries.
We can, in a similar fashion, construct more edge observables
by allowing the quantities analogous
to $S$ and $V_{a}$, denoted now by $T$ and $W_{a}$ respectively, to vary
as arbitrarily as possible at $\partial \Sigma$ such that (\ref{9.5}) and
(\ref{10}) turn out to be differentiable after adding
suitable
surface terms. They will then generate
good canonical
transformations and will be well-defined in the canonical framework.
Such observables are edge observables because they depend
only on the boundary values taken by the test
forms/functions (and their derivatives in the latter case)\cite{bal,regge}.
This can be seen
from
the fact that the difference of two of
these observables with different smearing forms/functions which coincide (along
with derivatives in the case of the latter) only
at the
boundaries is a constraint. This is as in the case of $H$. Also these
 observables are gauge invariant having weakly zero PB's with the constraints.
This  will be shown later.

We will first look
at the edge observables that come from the vector constraint.  To do this, we
first  rewrite the vector constraint in (\ref{9.5}) after a partial
integration as
\be
{\cal V}_{V} = \int _{\Sigma} d^{3}x\;q_{ab}{\cal L}_{V}p^{ab}.\label{redef}
\ee
In the above, let us replace $V$ by $W$ where $W$ is any vector field.
We require of $W$ that,
 at the boundaries of the manifold, it is
tangential to the boundary. Then it can be verified that the quantity so
obtained, namely
\be
{\cal D}_{W}=\int_{\Sigma}d^{3}x\;q_{ab}{\cal L}_{W}p^{ab}.\label{diff}
\ee
continues to be differentiable in both $q_{ab}$ and $p^{ab}$.
It furthermore has weakly zero PB's with the constraints :
\begin{eqnarray}
\{ {\cal D}_{W}\;,\;{\cal V}_{V}\} & = & {\cal V}_{[W\;,\;V]},
\nonumber \\
\{ {\cal D}_{W}\;,\;{\cal S}_{S}\} & = & {\cal S}_{{\cal L}_{W}S}.\label{12}
\end{eqnarray}
The right hand sides in these equations are constraints and hence
weakly zero because
their respective test fields are easily verified
to satisfy the conditions (\ref{bc1}) and (\ref{bc2}).

The algebra of observables generated by ${\cal D}_W$ is seen to be
\be
\{ {\cal D}_{W_{1}}\;,\;{\cal D}_{W_{2}}\} = {\cal D}_{[W_{1}\;,\;W_{2}]}.
\label{13}
\ee

We are interested in observables which are supported
at the edge corresponding to the event horizon rather than those which are
supported at spatial infinity. We will therefore hereafter assume that $W$ is
non-zero only at the
inner boundary and vanishes like $V$ at the boundary at infinity.

We next  define the observables which can be got from a partial
integration of the scalar constraint. Let us first look at the scalar
constraint
${\cal S}_S$ :
\be
{\cal S}_{S}=\int _{\Sigma}d^{3}x\;
S[-{q^{\frac{1}{2}}}
\;{^{(3)}R}+q^{-\frac{1}{2}}(p^{ab}p_{ab}-\frac{p^{2}}{2})].\label{14}
\ee
The above is clearly differentiable in $p^{ab}$.
As for
differentiability in $q_{ab}$, it can be verified that a variation of
$q_{ab}$ induces surface terms in its variation. They vanish only if
the test functions $S$ satisfy (\ref{bc2}).
The condition on their derivatives emerges
because variation of ${^{(3)}R}$ contains second derivatives of the variation
of the metric $q_{ab}$. The boundary condition  in
(\ref{bc2}) on $S$ are in fact
got from this
requirement of differentiability of ${\cal S}_S$.

Our task is to find  differentiable observables using formal partial
integration in (\ref{14}).  We proceed as follows.  Consider (\ref{14})
with $S$ replaced by $T$.    $T$
now does not have to satisfy the boundary
conditions satisfied by $S$.  We have already seen above that differentiability
in $q_{ab}$ of (\ref{14}) requires $S$ along with its derivatives to vanish at
the boundary.  Here we instead keep track of the variations to see if we can
cancel them with other surface terms.  The only term in this expression that
requires careful scrutiny is
\be
\label{curvat} \int _{\Sigma}d^{3}x\; T[-{q^{\frac{1}{2}}}\; {^{(3)}R}].\ee
The change in above term due to a variation $\delta q_{ab}$ is \cite{wald}
\be
-\int _{\Sigma}d^{3}x\; Tq^{\frac{1}{2}}[\frac{1}{2}\; {^{(3)}R}\;{q^{ab}}
- {^{(3)}
R^{ab}}]\delta q_{ab}-\int
_{\Sigma}d^{3}x\; T q^{\frac{1}{2}}[D^{a}D^{b}(\delta q_{ab})-D
^{a}(q^{cd}D_{a}\delta q_{cd})]. \label{varia}
\ee
Since the second term above contains derivatives of $\delta q_{ab}$,
(\ref{curvat}) is not differentiable with respect to $q_{ab}$.

Suppose  now that
\begin{eqnarray}
&&\delta q_{ab}|_{\partial \Sigma}=0, \label{(a)}\\
&&D_{a}T|_{\partial \Sigma}=0.
\label{(b)}
\end{eqnarray}
[ Note that (\ref{(b)}) implies that $T$ at the boundary
goes to a constant which can be non-zero.]
The terms involving derivatives of $\delta q_{ab}$ in
(\ref{varia}) give rise to surface terms in the variation. These surface
terms are now exactly cancelled by the
variation of
 \[ -2\int _{\partial \Sigma}T{\cal K}\sqrt{h} \]
where ${\cal K}_{ab}$ and $h_{ab}$ are respectively the extrinsic curvature and
the induced metric of the boundary $\partial \Sigma$ \cite{wald}.
Thus so long as the
conditions (\ref{(a)}) and (\ref{(b)}) above are met, we can define an edge
observable of the form
\be
{\cal H}_{T}=\int _{\Sigma}d^{3}x\; T[-{q^{\frac{1}{2}}}\;{^{(3)}
R} +q^{-\frac{1}{2}}(p^{ab}p_{ab}-
\frac{p^{2}}{2})]-2\int _{\partial \Sigma}d^{2}x\;
h^{\frac{1}{2}}TK.\label{edH}
\ee

Note that the Hamiltonian $H$ defined in (\ref{new}) is just
${\cal H}_{\Gamma}$:
\be
H = {\cal H}_{\Gamma}
\label{Hamil}
\ee
Thus $H$ is also an edge observable as indicated in the last
Section except that it is confined to the asymptotic boundary rather than the
inner boundary ( $\Gamma$ being non-vanishing only at spatial infinity ).

In contrast, $T$ goes to a constant at the boundary
surrounding
the hole.  When this boundary value is  non-zero, the observable
${\cal H}_{T}$ acts as a Hamiltonian which
non-trivially evolves observables at
the inner boundary.  Such an observable is physically
required by any stationary observer who is sufficiently close to the event
horizon (and
outside it) because it  will function as the time evolution
operator of this observer.

The interesting fact  is that all the other edge observables ${\cal D}_{W}$
defined earlier cease to be well-defined for this observer.  The reason is that
we required also the condition (\ref{(a)}) to make ${\cal H}_{T}$
well-defined,.
Such a condition is clearly incompatible with
the existence of the observables ${\cal D}_{W}$ because the latter generate
non-trivial diffeomorphisms at the boundary.  (In some special cases,
there may exist certain symmetries of the background metric so that the
boundary possesses Killing vector fields.  In those cases, we are
still permitted those edge observables that generate
diffeomorphisms along the Killing vector fields of the boundary.  However, it
is
nevertheless generically true that only a finite set out of the potentially
infinite set of edge observables survives as observables for this person.)

Conversely, for the asymptotic observer, there is no necessity for
introducing ${\cal H}_{T}$ as a well-defined observable.  In fact, this
observable is quite unphysical for this observer because
it performs ``time evolution'' at a boundary which from this person's
viewpoint has
``stopped'' evolving because of gravitational time dilation.  Thus, for this
observer, the infinite observables generating diffeomorphisms at the
inner boundary do exist as well-defined observables.

Such a difference between two types of stationary observers seems
to be similar to the black hole complementarity principle recently
envisaged by \cite{susskind}.  In \cite{susskind}, the complementarity is
between a stationary external observer and an in-falling observer, whereas here
the complementarity is between two stationary external observers, one of them
located at spatial infinity and the other located close to the event horizon.
The result found here may be significant
because it has arisen purely out of technical reasons having to do with
general requirements of compatibility of test function spaces suitable
for these observers.

Before we go on to the next Section, certain points about these edge
observables are worth noting.
One triviality to be noted is that they commute with $H$ ( $=$
${\cal H}_{\Gamma}$ in (\ref{Hamil})) as
mentioned in the introduction. Therefore the eigenstates of
$H$ will have an infinite
degeneracy corresponding to the edge states carrying the representation of
the diffeomorphism
algebra (\ref{13}).  One possibility to explore is whether this infinite
degeneracy of states for
any given energy level causes a ``particle'' to be in any one of these states
with
equal likelihood and whether  this randomness is what leads to the entropy
of the
black hole.  If such is the case, it would be very interesting to examine
 if the
logarithm of the degeneracy actually increases directly in proportion to the
area \cite{entropy}.  Clearly some
form of regularization is required in this calculation, since the degeneracy
is infinite to
begin with and  it does not make sense to talk about the dependence of
an infinite quantity on area. As this way of arriving at an
entropy does not
involve matter anywhere, it would be purely intrinsic to the gravitational
degrees of freedom.

\section{The Quantum Hall Effect}

As promised, we devote this section to  a well-studied condensed matter system.
Later on we will borrow the ideas used here to analyze the gravity systems.
All of what we
are going to describe below for the Hall system is well-understood by workers
in the field of quantum Hall effect (QHE) \cite{Prange}.

Classical Hall effect is the phenomenon of a longitudinal electric field
causing a transverse current in the presence of a perpendicular magnetic field.
A simple effective action that describes the physics is the Chern-Simons action
added on to the usual electromagnetic action:
\bea
S_{bulk}&=& \int_{\cal M} d^3x \left[ -\frac{1}{4e^2}F_{\mu \nu}F^{\mu \nu} -
\frac{\sigma_H}{2} \epsilon^{\mu \nu
\lambda}\,\,A_\mu \partial_\nu
A_\lambda \right] \label{MCS} \\
F_{\mu \nu}&=& \partial_\mu A_\nu - \partial_\nu A_\mu \nonumber
\eea
Here $e^2$ is a constant related to the inverse of the ``effective
thickness" of the Hall sample, while our metric is $(-1,+1,+1)_{diagonal}$.

The equation of motion of the above action is
\be
\frac{1}{e^2}\partial_\mu F^{\mu \nu} = \sigma_H \epsilon^{\nu \rho \lambda}
F_{\rho \lambda}
\label{eq}
\ee
Thus we see that the current $j^{\nu}$ for this theory is to be identified
with $\sigma _{H}\epsilon ^{\nu\rho\lambda}\partial _{\rho}A_{\lambda}$.  In
particular,
\be
j^i = -\sigma_H \epsilon^{ij}E_j
\label{heq}
\ee
Thus the $\sigma _{H}$ that appears as the coefficient of the Chern-Simons term
is the same as the Hall conductivity.

The connection of the above system with edge observables is also well-known.
The latter arise when we confine the above theory to a finite geometry (as is
appropriate for any
physical Hall sample).  From very general arguments first forwarded by Halperin
\cite{halperin}, the existence of chiral edge currents at the boundary can then
be established.

Now naively what happens is that the theory in the bulk
(described by (\ref{MCS})) does not communicate with the theory describing
these chiral currents at the edge. It is then not clear how these edge
currents can have any role in the description of bulk phenomena.

What saves the situation, however, is the fundamental requirement of
gauge invariance \cite{duality}. Thus the action (\ref{MCS}) under gauge
transformation $A \ra A + d\alpha$
changes by the surface term
\be
\frac{\sigma_H}{2}\int_{\partial {\cal M}} d\alpha \wedge A
\label{st}
\ee
But, physics is gauge invariant. Therefore  it must be that there is a theory
at the
boundary describing the chiral edge currents which  is also gauge
non-invariant such
that the total action ($S_{tot}=S_{bulk}+S_{edge}$) is itself gauge invariant.
This line of argument \cite{duality} then leads us to the  action
\be
S_{tot} = S_{CS} +\frac{\sigma_H}{2}
\int_{\partial {\cal M}} d\phi \wedge A -
\frac{\sigma_H}{4}
\int_{\partial {\cal M}} {\rm D}_\mu\phi\;{\rm D}^\mu \phi.  \label{totS}
\ee
\be
{\rm D}_\mu \phi = \partial_\mu\phi + A_\mu.  \nonumber
\ee

The field $\phi$ under a gauge transformation transforms as
\be
\phi \ra  \phi - \alpha.  \label{gtphi}
\ee
so that
\be
{\rm D}\phi \equiv d\phi + A \ra {\rm D}\phi \label{codep}
\ee
and
\be
S_{tot} \ra S_{tot}.
\label{gttot}
\ee
The second term in (\ref{totS}) is the term which restores gauge invariance.
Notice that it is the anomalous coupling alluded to in the Introduction.
The last term is a kinetic energy term, it is required if the
theory at the edge is to give rise to a {\em chiral} theory. [See
\cite{duality}
for details
on how this action can be used very fruitfully to obtain results in the FQHE.]

All of the above material is standard in the literature on the Hall effect.

Before dealing with Einstein gravity in (3+1) dimensions,
let us deal with a simple toy model in (2+1) dimensions in Section V
in order to get ideas fixed.

\section{How The Edge Couples To The Bulk In (2+1) Gravity}

As a toy model, we now show the explicit coupling  between the ``edge" and the
exterior in the case of 2+1 dimensional
gravity. The model considered below is related to the one considered in
\cite{al} and more recently in
Carlip \cite{carlip}. See also \cite{don}.

It is well-known \cite{witten} that the standard Einstein-Hilbert action on
a three-manifold admits a
reformulation as an $ISO(2,1)$-gauge theory with the Chern-Simons action
\be
S= \kappa \int_{\cal M} \tr (A \wedge dA + \frac{2}{3} A\wedge A \wedge A )
\label{action}
\ee
where
\be
A = e^a P_a + \omega^m J_m \equiv e\cdot P + \omega \cdot J.
\label{decomp}
\ee
and $\kappa$ is a constant which we hereafter set equal to one. We will
use the form and explicit index notations interchangeably. Here $A,
e^a,\omega^m$ are one-forms. Also $P_a$ and $J_m$ are the translation and
Lorentz group generators
respectively. They satisfy the ISO(2,1) Lie algebra relations

\bea
\left[P_a,P_b\right]&=&0,  \nonumber \\
\left[P_a, J_m\right] &=& \epsilon_{amb}P_b,  \nonumber  \\
\left[J_m,J_n\right] &=& \epsilon_{mnk}J_k.
\label{algebra}
\eea
Also, the trace appearing in (\ref{action}) is
given by,
\bea
\tr(P_aP_b)&=&0, \nonumber  \\
\tr ( P_a J_m)&=&\delta_{am}, \nonumber  \\
\tr (J_mJ_n)&=& 0.
\label{trace}
\eea

The action (\ref{action}) is not invariant under arbitrary gauge
transformations of the form
\be
A \ra  A'= uAu^{-1} + u\,du^{-1}.
\label{trans}
\ee
To remedy this situation, we introduce fields $g$ valued in the
$ISO(2,1)$ group
transforming under $u$ as
\be
g \ra ug
\label{gtrans}
\ee
We can now define the ``covariant"
derivative
\be
{\rm D}g \equiv dg + A g,
\label{defco}
\ee
which has the transformation property
\be
{\rm D}g \ra u{\rm D}g,
\label{tranco}
\ee
which allows us to construct the gauge invariant one-form
\be
{\cal A} \equiv g^{-1}Dg = g^{-1}dg + g^{-1}Ag
\label{invform}.
\ee

We can now construct the gauge invariant ``Chern-Simons"-like action $S_{inv}$
simply by
replacing $A$ by ${\cal A}$ in (\ref{action}) \cite{TRG} :
\be
S_{inv}({\cal A}) = \int_{\cal M} \tr\left[{\cal A}\wedge d{\cal A} +
\frac{2}{3} {\cal
A}\wedge {\cal A} \wedge {\cal A}\right].
\label{modaction}
\ee
It can be expanded, using (\ref{invform}), as follows :
\be
S_{inv} = S_{CS}(A) -\frac{1}{3}\int_{\cal M}\tr (g^{-1} dg)^3 -
\int_{\partial{\cal M}} \tr (dg g^{-1}\wedge A).
\label{fullaction}
\ee
Here the first term is the usual Chern-Simons action, the second term is the
Wess-Zumino-Novikov-Witten (WZNW) term and the last term gives the anomalous
coupling between
the variable $g$ and the gauge field $A$. We have also assumed that ${\cal M}$
has boundary $\partial {\cal M}$, arising from the boundaries of its spatial
cuts \cite{TRG}.

The form of such a gauge invariant action (\ref{fullaction}) is the same
for arbitrary $G$. We  now specialize it to the  group $ISO(2,1)$. Let $H
=\{h\}$ be its subgroup of translations and let $\{w\}$ denote its $SO(2,1)$
subgroup. Any $g \in
\; ISO(2,1)$ can then be written as
\bea
g = hw = e^{\chi} w,  \nonumber \\
h \equiv e^\chi.   \label{abelian}
\eea
Using the fact that
$h^{-1}dh \equiv d \chi$, we
can now reduce the WZNW term to
\bea
\int_{\cal M} \tr (g^{-1}dg)^3  &=& \int_{\cal M} \left[
3 \tr (d \chi \wedge d(dww^{-1}) + \tr (dw w^{-1})^3 \right] \nonumber \\
&=& \int_{\cal M}\left[ -3 \tr d( d\chi \wedge dw w^{-1}) \right] \nonumber \\
&=& - 3 \int_{\partial{\cal M}} \tr ( d \chi \wedge d w w^{-1}).
\label{reduce}
\eea
Here we have used  (\ref{trace}) to get rid of the term
involving $w$ only.

The gauge invariant action now reduces to
\be
S_{inv}= S_{CS}(A) + \int_{\partial{\cal M}} \tr
\left[ d\chi \wedge (dw w^{-1}
- A) - h dw w^{-1}h^{-1} \wedge A \right].
\label{nonkinetic}
\ee
This equation  shows that the variable  $g$ couples to the gauge field
$A$ only at the edge, just as in QHE. However, as it stands,
the field $g$  does not have a dynamics of its
own. We therefore add a
gauge-invariant kinetic energy term for the $g$ field  to the action
living only on the
boundary $\partial {\cal M}$:
\be
S_{kin} = \frac{1}{2} \int_{\partial{\cal M}} \, d^2x \, {\sqrt{ det \,\eta}}
\eta^{\mu \nu}
\tr \left[ {\rm D}_{\mu}g^{-1} {\rm D}_{\nu} g \right].
\label{kin}
\ee
Here $ {\rm D}_{\mu}g = \partial_{\mu}g + A_\mu g $  is the covariant
derivative. Also, $\eta^{\mu \nu}$ is the metric on $\partial {\cal M}$
induced by
the metric on ${\cal M}$.

Note that (\ref{kin}) is the natural analogue of the kinetic energy term in
(\ref{totS}) for QHE.

Using (\ref{abelian}) and (\ref{trace}), we can expand the action
(\ref{kin}) as follows :
\bea
S_{kin} = - \int _{\partial {\cal M}} \,d^2x {\sqrt{ det \eta}}
\;\eta ^{\mu \nu } &\tr& [ \partial_{\mu}
\chi \left( \partial_\nu w w^{-1} + \frac{1}{2}(J \cdot \omega _\nu
+\frac{1}{2} w^{-1} A_\nu w) \right)
 + \nonumber \\
&e_\mu & \omega_\nu
+ \frac{1}{2} \left( e_\mu \cdot P (w^{-1}\partial_\nu w +
\partial_\nu w w^{-1} +  \partial_\mu w w^{-1} [ \omega_\nu \cdot J,
\chi ] \right) ]
\label{expand}
\eea

The total action is the sum of (\ref{nonkinetic}) and (\ref{expand}). Notice
that the dynamics of the Chern-Simons field $A$ is completely determined in
${\cal M}$ while the fields $\chi$ and $w$ get their dynamics on the edge.
Also they couple to $A$ via the anomalous term in (\ref{fullaction}) and
the kinetic energy term (\ref{expand}). The edge variables thus couple to the
bulk variables.

\section{Final Remarks: The Edge/Bulk Connection in (3+1) Gravity}

We now attempt to repeat these arguments for the case of  (3+1) gravity.
We first observe that we have an analogue of (\ref{MCS}) in the bulk (namely,
the exterior of the black hole) which is the Einstein-Hilbert action
(\ref{EH}).  We also have
an analogue
of the edge currents which are just the edge observables defined in
(\ref{diff}). Note that we now imagine ourselves to be
in the asymptotic frame of
reference
(rather than in the frame of reference of an observer close to the horizon) for
specificity so
that we do have the infinite set of edge observables corresponding to
spatial diffeomorphisms of the horizon.

There is also a notion of ``gauge invariance'' which here
is the general coordinate
or diffeomorphism invariance.  Thus we have all the necessary ingredients that
the Hall system had.  We need to check just one last feature: namely, does the
Einstein Hilbert action (\ref{EH}) suffer from diffeomorphism non-invariance?
If it does, and if we can postulate an anomalous
coupling to restore diffeomorphism invariance, then the analogy would be
complete.  This coupling  will serve as the
crucial missing link connecting the exterior and the
surface of the black hole.

It is very interesting therefore the action (\ref{EH}) is not diffeomorphism
invariant for a manifold with a boundary.  The reason of course is very
trivial.
If we consider the Einstein-Hilbert action restricted to the exterior of the
black hole and perform {\it arbitrary} diffeomorphisms, then
\be
\delta _{v}\int _{\Sigma \times \real }d^{4}x\;\sqrt{-g}\; ^{4}R
= -\int _{\partial \Sigma \times \real }d^{3}x\;\sqrt{-f}v^{a}u_{a}\;
^{(4)}R. \label{aneh}
\ee
Here $v^{a}$ is the vector field generating the above diffeomorphism,
$\delta_v$ induces the corresponding variations on fields, $f_{ab}$
is the induced metric at the boundary and $u_{a}$ is the unit normal at the
boundary. [ See also \cite{don}. ]

If $v^{a}$ is tangential to the boundary, then the above variation
(\ref{aneh}) is zero and then the action is unchanged.
But there does not seem to be any reason to constrain $v^{a}$ to be tangential
to the boundary (at the boundary) especially because we know that the full
theory (which includes also the interior) is diffeomorphism invariant under all
possible diffeomorphisms.  The above variation is non-zero for such
diffeomorphisms.  So we do need an anomalous coupling just as in the case of
the Hall effect.

At this point we neither have an explicit action describing the dynamics at the
edge nor the ``anomalous'' term which provides the coupling, but the essential
point is that a fully diffeomorphism invariant effective theory {\em must}
include these surface terms.  These ideas suggest that any description about
loss of information which looks purely at the states in the bulk is necessarily
incomplete because there is also the surface action (and associated edge
states) to contend with.  Just as diffeomorphism non-invariance may be
avoided
by the presence of surface terms, it seems reasonable to suppose that
information loss too 
may be avoided or mitigated by the presence of such terms.

\noindent
\section*{Acknowledgements}
It is a great pleasure to thank Rafael Sorkin for valuable comments.
The derivation of (\ref{fullaction}) from (\ref{action}) was accomplished,
in a
related context, in collaboration with T. R. Govindarajan, to
whom we are therefore especially thankful.
We are also grateful to  E. Ercolessi, G. Jungman,
B. Sathiapalan and J. Varghese for discussions.  This work was supported
by  US DOE contract number DE-FG02-85ER40231 and
by a Syracuse University Graduate fellowship awarded to A.M.

\end{document}